\documentclass[a4paper,12pt]{article}
\usepackage{epsfig}
\usepackage{amssymb,latexsym,amsfonts,amsmath}
\usepackage{graphicx}

\newtheorem{conjecture}{Conjecture}

\newskip\humongous \humongous=0pt plus 1000pt minus 1000pt

\newif\ifdtup

\catcode`\@=11

\@addtoreset{equation}{section}
\def\theequation{\thesection.\arabic{equation}}

\def\@normalsize{\@setsize\normalsize{15pt}\xiipt\@xiipt
\abovedisplayskip 14pt plus3pt minus3pt%
\belowdisplayskip \abovedisplayskip
\abovedisplayshortskip \z@ plus3pt%
\belowdisplayshortskip 7pt plus3.5pt minus0pt}

\def\small{\@setsize\small{13.6pt}\xipt\@xipt
\abovedisplayskip 13pt plus3pt minus3pt%
\belowdisplayskip \abovedisplayskip
\abovedisplayshortskip \z@ plus3pt%
\belowdisplayshortskip 7pt plus3.5pt minus0pt
\def\@listi{\parsep 4.5pt plus 2pt minus 1pt
      \itemsep \parsep
      \topsep 9pt plus 3pt minus 3pt}}

\relax

\catcode`@=12

\catcode`\@=11

\def\section{\@startsection{section}{1}{\z@}{3.5ex plus 1ex minus
    .2ex}{2.3ex plus .2ex}{\large\bf}}

\def\thesection{\arabic{section}}
\def\thesubsection{\arabic{section}.\arabic{subsection}}

\def\appendix{\setcounter{section}{0}
  \def\thesection{Appendix \Alph{section}}
  \def\thesubsection{\Alph{section}.\arabic{subsection}}
  \def\theequation{\Alph{section}.\arabic{equation}}}

\def\SymBoxes#1#2#3#4{\newdimen\un@t \un@t#3%
\raisebox{#1}{\rule{#2\un@t}{#4}\hskip-#2\un@t% lower horizontal
\@tempdimb\un@t \advance\@tempdimb by-#4\@tempcntb#2\relax%
\@whilenum{\@tempcntb>0}\do{%                         % #2 vertical lines
\rule{#4}{\un@t}\hskip\@tempdimb \advance\@tempcntb by\m@ne}%
\hskip-#2\un@t \rule[\un@t]{#2\un@t}{#4}%
\rule[\un@t]{#4}{#4}\hskip-#4%             % upper horizontal line
\rule{#4}{\un@t}}\hskip-#4}                % rightest vertical line

\begin{document}

\newcommand{\dd}{\textrm{d}}

\newcommand{\beq}{\begin{equation}}
\newcommand{\eeq}{\end{equation}}
\newcommand{\bea}{\begin{eqnarray}}
\newcommand{\eea}{\end{eqnarray}}
\newcommand{\beas}{\begin{eqnarray*}}
\newcommand{\eeas}{\end{eqnarray*}}
\newcommand{\defi}{\stackrel{\rm def}{=}}
\newcommand{\non}{\nonumber\\}
\newcommand{\bquo}{\begin{quote}}
\newcommand{\enqu}{\end{quote}}
%%%%%%%%%%%%%%%%
\renewcommand{\(}{\begin{equation}}
\renewcommand{\)}{\end{equation}}
%%%%%%%%%%%%%%%%%%%%%%%%%%%%%%%%%% definitions
\def\de{\partial}
\def\Om{\ensuremath{\Omega}}
\def\Tr{ \hbox{\rm Tr}}
\def\rc{ \hbox{$r_{\rm c}$}}
\def\H{ \hbox{\rm H}}
\def\HE{ \hbox{$\rm H^{even}$}}
\def\HO{ \hbox{$\rm H^{odd}$}}
\def\HEO{ \hbox{$\rm H^{even/odd}$}}
\def\HOE{ \hbox{$\rm H^{odd/even}$}}
\def\HHO{ \hbox{$\rm H_H^{odd}$}}
\def\HHEO{ \hbox{$\rm H_H^{even/odd}$}}
\def\HHOE{ \hbox{$\rm H_H^{odd/even}$}}
\def\K{ \hbox{\rm K}}
\def\Im{ \hbox{\rm Im}}
\def\Ker{ \hbox{\rm Ker}}
\def\const{\hbox {\rm const.}}
\def\o{\over}
\def\im{\hbox{\rm Im}}
\def\re{\hbox{\rm Re}}
\def\bra{\langle}\def\ket{\rangle}
\def\Arg{\hbox {\rm Arg}}
\def\exo{\hbox {\rm exp}}
\def\diag{\hbox{\rm diag}}
\def\longvert{{\rule[-2mm]{0.1mm}{7mm}}\,}
\def\a{\alpha}
\def\b{\beta}
\def\e{\epsilon}
\def\l{\lambda}
\def\ol{{\overline{\lambda}}}
\def\ochi{{\overline{\chi}}}
\def\th{\theta}
\def\s{\sigma}
\def\oth{\overline{\theta}}
\def\ad{{\dot{\alpha}}}
\def\bd{{\dot{\beta}}}
\def\oD{\overline{D}}
\def\opsi{\overline{\psi}}
\def\dag{{}^{\dagger}}
\def\tq{{\widetilde q}}
\def\L{{\mathcal{L}}}
\def\p{{}^{\prime}}
\def\W{W}
\def\N{{\cal N}}
\def\hsp{\ ,\hspace{.7cm}}
\def\bo{\ensuremath{\hat{b}_1}}
\def\bfo{\ensuremath{\hat{b}_4}}
\def\co{\ensuremath{\hat{c}_1}}
\def\cfo{\ensuremath{\hat{c}_4}}
\newcommand{\C}{\ensuremath{\mathbb C}}
\newcommand{\Z}{\ensuremath{\mathbb Z}}
\newcommand{\R}{\ensuremath{\mathbb R}}
\newcommand{\rp}{\ensuremath{\mathbb {RP}}}
\newcommand{\cp}{\ensuremath{\mathbb {CP}}}
\newcommand{\vac}{\ensuremath{|0\rangle}}
\newcommand{\vact}{\ensuremath{|00\rangle}                    }
\newcommand{\oc}{\ensuremath{\overline{c}}}

\newcommand{\Vol}{\textrm{Vol}}

\newcommand{\half}{\frac{1}{2}}

\begin{titlepage}
\def\thefootnote{\fnsymbol{footnote}}

\begin{center}
{\large {\bf High $Q$ BPS Monopole Bags are Urchins }}
\end{center}

\bigskip
\begin{center}
{\large Jarah Evslin\footnote{\texttt{jarah(at)ihep.ac.cn}} 
and Sven Bjarke Gudnason\footnote{\texttt{gudnason(at)phys.huji.ac.il}}}
\end{center}

\renewcommand{\thefootnote}{\arabic{footnote}}

\begin{center}
\vspace{0em}
{\em {TPCSF, Institute of High Energy Physics, CAS, P.O.~Box 918-4,
    Beijing 100049, P.R.~China\\and\\ 
Racah Institute of Physics, The Hebrew University, Jerusalem 91904, Israel
\vskip .4cm}}

\end{center}

\vspace{1.1cm}

\noindent
\begin{center} {\bf Abstract} \end{center}

\noindent
It has been known for 30 years that 't Hooft-Polyakov monopoles of
charge $Q$ greater than one cannot be spherically symmetric.  5 years
ago, Bolognesi conjectured that, at some point in their moduli space,
BPS monopoles can become approximately spherically symmetric in the
high $Q$ limit.  In this note we determine the sense in which this
conjecture is correct.  We consider an $SU(2)$ gauge theory with an
adjoint scalar field, and numerically find configurations with $Q$
units of magnetic charge and a mass which is roughly linear in $Q$,
for example in the case $Q=81$ we present a configuration whose energy
exceeds the BPS bound by about $54$ percent.  These approximate
solutions are constructed by gluing together $Q$ cones, each of which
contains a single unit of magnetic charge. 
In each cone, the energy is largest in the core, and so a constant
energy density surface contains $Q$ peaks and thus resembles a
sea urchin.   We comment on some relations between a non-BPS cousin of these solutions and  the dark matter halos of dwarf spherical galaxies.

\vfill

\begin{flushleft}
{\today}
\end{flushleft}
\end{titlepage}

\hfill{}

\setcounter{footnote}{0}

\section{Introduction}

Within the large moduli space of solutions of BPS monopoles
\cite{Ward:1981jb,Prasad:1980hg,Prasad:1980hi,Prasad:1980yw} of charge
$Q$, it is plausible that there exists a high $Q$ limiting 
sequence of monopoles that are spherically symmetric up to $1/Q$
corrections. However even the most qualitative features of such
solutions are, as yet, unknown.   
In this note we will solve an easier problem, we will explicitly 
construct configurations in an $SU(2)$ gauge theory with an adjoint 
scalar whose energy slightly exceeds the BPS bound.  These
configurations do not provide time-independent solutions of the
equations of motion\footnote{They do provide initial conditions for
  oscillatory time-dependent monopole solutions.}, but the near
saturation of the BPS bound supports a conjecture 
that certain qualitative features of these solutions will be shared by 
some true solutions.  In particular, our approximate solutions become  
spherically symmetric in the high $Q$ limit, and so the corresponding
true solutions are also asymptotically spherically symmetric. 

We will construct these approximate solutions by decomposing space
into $Q$ identical cones which extend from the origin, each of which
asymptotically contains a unit of magnetic flux and is axially
symmetric in a sense which will be made precise below.  Of course, the
classification of regular polyhedra implies that no decomposition of
space into identical cones exists for general $Q$.  Therefore the
cones may not fill all of space.  We will provide configurations in
which the space 
between the cones consists of a vanishing gauge field and a continuous
Higgs field, which at high $Q$ provide a 
contribution to the energy which decreases with $Q$, and so is
subdominant with respect to the total energy of the configurations,
which is proportional to $Q$.

Our problem is then reduced to the following steps.  First, in 
Sec.~\ref{ansez} we will choose an Ansatz and boundary conditions for 
the configuration in a single cone.  The finiteness of the energy
fixes all of the functions in our Ansatz except for two angular
functions which must be chosen.  Different choices yield different
energies, none of which satisfy the BPS bound but all of which satisfy
the bound up to a factor which is $Q$-independent at large $Q$. 
This $Q$-independence, which is critical for the qualitative agreement
with a BPS configuration, is achieved if we impose that the radial
dependence of the solution satisfies an ODE in a single variable. Next
in Sec.~\ref{qualsez} we will explicitly solve this equation in 
the large and small radius limits, and discuss how these solutions
grow together.  In this way we are able to establish the 
qualitative profiles of these solutions, confirming the monopole bag
description conjectured in Ref.~\cite{Bolognesi:2005rk}.  We will
then present numerical solutions of the ODE in Sec.~\ref{numsez}.
Given the large moduli space of solutions, one may wonder why
approximately spherically symmetric monopoles are interesting,
especially given the fact that exact solutions which are spherically
symmetric up to $1/Q$ corrections are unknown.  In Sec.~\ref{scurasez}
we describe the kinds of field theories in which we expect such
monopoles to be the only monopoles which survive a non-BPS
deformation, and provide a very speculative application of such
theories. 

Since Bolognesi's groundbreaking proposal \cite{Bolognesi:2005rk}, the field of monopole bags has been steadily advancing with the understanding of various limiting behaviors \cite{Lee:2008ze} and of large $Q$ limits of Nahm transforms of bag solutions \cite{Harland:2011tm}.  Yesterday a paper by Manton appeared on the arXiv with 
significant overlap with our results \cite{Manton:2011vm}.   He found
exactly BPS Bolognesi bag solutions with an approximate spherical
symmetry of the type considered here.  This begs the question: how we
can justify presenting an approximate solution after exact solutions
have appeared?  Our justification is that the physical application
that we have in mind in Sec.~\ref{scurasez} is in a theory in which we
expect the radial dependence to be very distorted, but under a radial
deformation the cone structure of our configurations is preserved.

\section{Ansatz and equations of motion} \label{ansez}

\subsection{The cone}

As described in the introduction, charge $Q$ magnetic monopole
configurations will be constructed from $Q$ identical cones extending 
from the origin together with an extrapolation in the region between
the cones which yields a contribution to the energy which is
subdominant at large $Q$. The crucial step in the
construction is therefore to provide the configuration in a single cone.
The first step is to determine the size of the cone and to fit it with
coordinates. 

For simplicity let $Q=n^2$ be a perfect square.  This limitation
will not be relevant for our analysis of the large $Q$ regime.  We will consider a cone whose core extends along the positive $z$
axis, and will let $\psi$ be the azimuthal angle 
\beq
\tan\psi = \frac{y}{x} \ .
\eeq
If we define the radius in cylindrical
coordinates to be
\beq
\rho \equiv \sqrt{x^2+y^2} \ , 
\eeq
then the cone will be defined by the condition
\beq
\rho \leq \sigma\frac{z}{n} \ , 
\eeq
where  the cone is small enough that $n^2$ 
non-overlapping copies fit inside of $\R^3$. The constant
$\sigma$ parametrizes the size of the cone.

The base of each cone at a radius $r=\sqrt{x^2+y^2+z^2}$ is then a
circle of radius  
$\rho=r/\sqrt{\frac{n^2}{\sigma^2}+1}\simeq \sigma r/n$ and so the
area of the base of the cone is about $\pi \sigma^2 r^2/n^2=\pi \sigma^2 r^2/Q$ which is $\sigma^2/4Q$ times
the total area at that radius, implying that $\sigma\leq 2$ in
order for the area of the cones to be less or equal to the area of the
spatial sphere.
This is still not sufficient for the $Q$ cones to fit inside of
$\R^3$. Here we are interested in the large $Q$ limit and hence we can
consider the size of the base of the cone to be much much smaller than
the radius of curvature of the sphere at $r$. Then the limit, which is
an upper bound on the size of the cone, is given by the size of the
circle which can fit inside of a hexagon\footnote{This hexagonal lattice approximation is exact for monopoles in AdS \cite{stefanop,sutcliffe}.}. More precisely, in the large $Q$ limit the upper bound on $\sigma$ is
\beq
\sigma_{\rm max} = 2\sqrt{\frac{A_{\rm circle}}{A_{\rm hexagon}}} =
2\sqrt{\frac{\pi}{2\sqrt{3}}} \simeq 1.9 \ .
\label{eq:sigmamax}
\eeq

Clearly such a configuration cannot hope to saturate the BPS bound,
instead a qualitative agreement with a true solution leads to the
requirement that it exceeds the BPS bound by a factor (of order unity)
which tends to a constant in the high $Q$ limit.

\subsection{The Ansatz}

We will consider an $SU(2)$ gauge theory with a massless adjoint
scalar $\Phi$. Furthermore, we will make the crude approximation that
the configuration of the Higgs and gauge field factorizes into $z$
and $\rho/z$-dependent functions, yielding the following
axially-symmetric Ansatz  
\beq
\Phi = h(z)\left[
F(\eta)\left(c\, t^1 + s\, t^2\right)
+\epsilon\sqrt{1-F^2(\eta)} \, t^3\right] \ , 
\label{eq:phi} 
\eeq
for the Higgs field and
\begin{align}
A_1 &=
\frac{\alpha(z)}{z}
\left(
cs\left[J(\eta) 
  - G(\eta)\right]t^1
+\left[c^2 G(\eta) + s^2 J(\eta)\right] t^2 
  - s H(\eta)\, t^3\right) \ , \nonumber\\
A_2 &= \frac{\alpha(z)}{z}
\left(-\left[c^2 J(\eta) + s^2 G(\eta)\right] t^1 
  -cs\left[J(\eta) - G(\eta)\right] t^2
  + c H(\eta)\, t^3\right) \ , \nonumber\\ 
A_3 &= \frac{\alpha(z)}{n z}
I(\eta)\left(s\, t^1 - c\, t^2\right) \ , \label{eq:A}
\end{align}
for the $SU(2)$ gauge field $A_i$ where we have defined the variables 
\beq
\eta \equiv \frac{n\rho}{z}\in[0,\sigma] \ , \quad
\epsilon \equiv {\rm sign}(F'(\eta)) \ , \quad
c \equiv \cos\psi \ , \quad
s \equiv \sin\psi \ ,
\eeq
and the functions $F$, $G$, $H$, $I$ and $J$ of $\eta$ and also $h$
and $\alpha$ of $z$. 

Such solutions are invariant under a rigid axial symmetry which
simultaneously rotates the vectors 
\beq
\begin{pmatrix}
x & t^1\\
y & t^2
\end{pmatrix} \rightarrow
\begin{pmatrix}
\cos\phi & -\sin\phi\\
\sin\phi & \cos\phi
\end{pmatrix}
\begin{pmatrix}
x & t^1\\
y & t^2
\end{pmatrix} \ , \label{vrot}
\eeq
and the potential
\beq
\begin{pmatrix}
A_1\\
A_2
\end{pmatrix} \rightarrow
\begin{pmatrix}
\cos\phi & \sin\phi\\
-\sin\phi & \cos\phi
\end{pmatrix}
\begin{pmatrix}
A_1\\
A_2
\end{pmatrix} \ , \label{arot}
\eeq
by an arbitrary angle $\phi$.

\subsection{Finite energy conditions}\label{sec:finiteenergyconditions}

\subsubsection*{Topological conditions}

We define the charge of the magnetic monopole to be equal to the
number of cones $Q=n^2$ in which the Abelian magnetic field $\Tr(F\Phi)$
asymptotically pointing in the outward radial direction integrates,
over the base of each cone, to a single Dirac quantum $4\pi$.  The
configuration then will have a finite energy only if, for all
sufficiently high $z$, the value of $\Phi$ on the cone's base is in
the correct topological sector.  More precisely, we will demand that,
at large $z$, $\Phi$ be constant on the boundary $\rho=\sigma z/n$ of the cone. 
Therefore the value of $\Phi$ on the base of the cone defines a map
from a 2-sphere $S^2$, which is the $(z=\infty,\ \eta\leq\sigma)$ two-disc $D^2$ with its boundary 
identified with a point, to the space $S^2$ of values of $\Phi$ of constant norm $|\Phi|=v$.  The topological condition is then that this
map $S^2\rightarrow S^2$ be of degree one.

We will satisfy this condition by imposing that $F$ be a continuous 
function such that 
\beq
0\leq F(\eta)\leq 1\hsp
F(0) = 0 \ , \qquad
F(\sigma) = 0 \ ,
\label{eq:BC1}
\eeq
and $F(\eta_0)=1$ for some intermediate value $\eta_0$ of its
argument. We will assume that the function $F$ has only one maximum on
the interval $[0,\sigma]$. 
One function obeying this condition is
\beq
F(\eta) = \sin\left(\frac{\pi\eta}{\sigma}\right) \ .
\label{eq:Fsine} 
\eeq
We will see shortly that the analysis of the region between the cones is simplified if one requires the derivative of $F$ to
be zero at $\eta=\sigma$ which can be achieved via the deformation
\beq
F(\eta) = \sin\left(\frac{\pi\eta}{\sigma}\right)
  \left(1-\frac{\eta^\kappa}{\sigma^\kappa}\right) \ ,
\label{eq:Fmodsine}
\eeq
where $\kappa\gg 1$ is a large integer. With this deformation, $F$ needs to be rescaled so that its maximal value is equal to one.  It is difficult to calculate the resulting normalization of $F$ analytically, but for $\kappa\gg 1$ the normalization constant is
arbitrarily close to unity.

\subsubsection*{High $z$ asymptotics}

The topological condition on the Higgs field is necessary but not
sufficient for the finite energy of the configuration.  The total
energy is the sum of the integrals of the gauge field strength energy
density $|F_{ij}|^2$ and the kinetic energy density of the Higgs field
$|D_i\Phi|^2$.  So long as the local energy density is finite, a
divergence may only arise from the noncompact region
$z\rightarrow\infty$, where we will demand that 
\beq
h(\infty) = g v \ , \qquad \alpha(\infty) = n \ , \label{eq:lim}
\eeq
and so their derivatives tend to zero.  The former condition is an
arbitrary choice, however in the non-BPS generalizations discussed in
Sec.~\ref{scurasez} it minimizes the potential for the Higgs field.
As will be clear from the analysis that follows, any different choice
of limiting value for $\alpha$ will lead to a rescaled value of the
finite energy conditions on the functions $G$, $H$, $I$ and $J$.  The
combination of the rescaling of $\alpha$ with that of the functions
leads to precisely the same form of the connection, and so this reflects
a simple redundancy in the parametrization of our Ansatz.

\subsubsection*{Higgs field kinetic energy}

We will now impose that the total energy in each cone is finite.
There are two contributions to the energy, one from the magnetic field 
$F_{ij}F^{ij}$ and another from the kinetic term of the Higgs field
$|D_i\Phi|^2$.  As both contributions are positive definite, we
must demand that they are finite separately.  We now begin by imposing
the finiteness of the Higgs field kinetic energy. 

The finiteness of the Higgs field kinetic energy
places non-trivial conditions on the functions in our Ansatz.  At high
$z$ the Higgs field tends to a constant limit, and so its covariant
derivative is at most of order $\mathcal{O}(1/z)$ and the energy
density $|D_i\Phi|^2$ is therefore at most of order
$\mathcal{O}(1/z^2)$. The finiteness of the energy of the
configuration is then equivalent to the vanishing of the order
$\mathcal{O}(1/z^2)$ and $\mathcal{O}(1/z^3)$ terms of $|D_i\Phi|^2$
for each spatial index $i$ and each gauge direction.  As there are
three spatial directions and three gauge directions, this consists of
$9$ directions, of which we will see $3$ are independent.   

The axial symmetry (\ref{vrot}) and (\ref{arot}) implies that the
energy density is axially symmetric.  Therefore it suffices to
calculate the energy at $\psi=0$, corresponding to $x=\rho$ and
$y=0$. This is merely to simplify the expressions in the following.
The three contributions to the kinetic energy are then the squares of
the quantities $D_i\Phi=\partial_i\Phi+i[A_i,\Phi]$,
\bea
D_x\Phi|_{y=0} &=& \frac{\epsilon h}{z}
\big(n|F'|-\alpha G\sqrt{1-F^2}\big)
\left(t^1-\frac{\epsilon F}{\sqrt{1-F^2}} \, t^3\right) \ , \non
D_y\Phi|_{y=0} &=& \frac{h F}{z}
\left(\frac{n}{\eta}-\alpha\left(\epsilon J
\frac{\sqrt{1-F^2}}{F}+H\right)\right)t^2 \ , \label{dcrudo} \\
D_z\Phi|_{y=0} &=& h'\big(F \, t^1+\epsilon\sqrt{1-F^2} \, t^3\big)
+\frac{\epsilon h}{z}\left(-\eta|F'|+\frac{\alpha}{n}\sqrt{1-F^2} I\right)
\left(t^1-\frac{\epsilon F}{\sqrt{1-F^2}} \, t^3\right) \ . \nonumber
\eea
The terms of order $\mathcal{O}(1/z^2)$ and $\mathcal{O}(1/z^3)$ in
the kinetic energy will only arise if there are terms of order
$\mathcal{O}(1/z)$ in $D_i\Phi$.  Therefore if we assume that $h-gv$
and $\alpha-n$ go to zero at least as quickly as $1/z$, then we may
drop the first term in the expression for $D_z\Phi$ and approximate
$\alpha$ by $n$ everywhere in (\ref{dcrudo}) without affecting the
divergent terms. 

The $\mathcal{O}(1/z)$ terms in each of the three $D_i\Phi$ can then
be seen to be proportional to a combination of the functions $F$, $G$,
$H$, $I$ and $J$.  By setting these combinations to zero, we eliminate
the divergence.   These three conditions can be solved to yield, for
example, $G$, $H$ and $I$ as functions of $F$ and $J$, we find
respectively 
\beq
G = \frac{|F'|}{\sqrt{1-F^2}} \ , \qquad 
H = \frac{1}{\eta}-\epsilon J\frac{\sqrt{1-F^2}}{F} \ , \qquad 
I = \frac{\eta|F'|}{\sqrt{1-F^2}} \ . \label{GHI}
\eeq
Recall that $F(\sigma)=0$ therefore $H$ is only finite if
$J(\sigma)=0$ as well, and in fact the ratio of the two needs to tend
to a constant as $z$ tends to $\sigma$.  $F(0)=0$ as well, and so the
finiteness of $H(0)$ requires that the divergences in both terms
cancel.  Here $\epsilon=1$ and so 
\beq
\lim_{\eta\to 0} \frac{J}{F} = 
\lim_{\eta\to 0} \frac{1}{\eta} + \mathcal{O}(1) \ . \label{JsuF}
\eeq

Substituting these relations into (\ref{dcrudo}) the $J$ dependence
cancels and so we can find exact expressions for the covariant
derivatives as functionals of $F$ alone 
\bea
D_x\Phi|_{y=0} &=& \frac{h F'}{z} (n-\alpha)
\left(t^1-\frac{\epsilon F}{\sqrt{1-F^2}}\, t^3\right) \ , \non
D_y\Phi|_{y=0} &=& \frac{h F}{z\eta} (n-\alpha) \, t^2
\ , \label{dcotto} \\
D_z\Phi|_{y=0} &=& h'\big(F \, t^1 + \epsilon\sqrt{1-F^2} \, t^3\big) 
- \frac{\eta}{n}D_x\Phi|_{y=0} \ . \nonumber
\eea
Notice that finite energy densities require
\beq
\lim_{z\to 0} \frac{h}{z} < \infty \ , \qquad 
\lim_{\eta\to 0} \frac{F}{\eta} < \infty \ , 
\eeq
and imply in particular
\beq h(0)=\alpha(0)=0 \ , \eeq
as they must be in order for $\Phi$ and $A_i$ to be well defined at the
origin.  A finite energy density at the maximum of $F$ also requires
that $F'$ vanish as quickly as $\sqrt{1-F^2}$. 

The total Higgs kinetic energy density is found by adding the squares of the components in Eq.~(\ref{dcotto}) 
\beq
\rho_{\rm kin} = \frac{1}{g^2}\sum_{i=1}^3 \Tr |D_i\Phi|^2 = 
\frac{1}{2 g^2} (h')^2
+\frac{h^2}{2 g^2 z^2}(n-\alpha)^2 \left(\frac{F^2}{\eta^2}
+\frac{(F')^2}{1-F^2}\left(1+\frac{\eta^2}{n^2}\right)\right) \ .
\label{eq:rhokin}
\eeq
As the covariant derivatives were independent of $J$, so is the total
kinetic energy density.

\subsubsection*{Magnetic field kinetic energy}

As was the case for the Higgs kinetic term, the axial symmetry of the
energy means that, for the purposes of calculating energy, it suffices
to consider the gauge field strength 
\beq
F_{ij} = \partial_i A_j-\partial_j A_i + i[A_i,A_j] \ ,
\eeq
at $y=0$.  Again this contribution to the total energy will be finite
if the energy density is at most of order $\mathcal{O}(1/z^4)$, which
means that the field strength itself must be at most of order
$\mathcal{O}(1/z^2)$.  Therefore it will again suffice to impose that
terms of order $\mathcal{O}(1/z)$ vanish, and that the coefficient of
the order $\mathcal{O}(1/z^2)$ terms is finite. 

These conditions are only nontrivial in the case of $F_{12}$, which is 
\beq
F_{12}|_{y=0} = \frac{\alpha}{z^2}
\left(\left[n\left(-J'+\frac{G-J}{\eta}\right)-\alpha G H\right]t^1
+\left[n\left(H'+\frac{H}{\eta}\right)-\alpha G J\right]t^3\right)
\ . \label{f12} 
\eeq
A divergence at the tip of the cone can be avoided if 
\beq
\lim_{z\to 0}\frac{\alpha}{z^2} < \infty \ . \label{alphazz}
\eeq
The expression (\ref{f12}) is then, at any constant $\eta$, of order
$\mathcal{O}(1/z^2)$ and so the energy is only divergent if the field
strength itself diverges.  To avoid such a divergence as $\eta$ tends
to $0$ we will impose that the two quotients by $\eta$ are finite 
\beq
\lim_{\eta\to 0} \frac{G-J}{\eta} < \infty \ , \qquad
\lim_{\eta\to 0} \frac{H}{\eta} = \lim_{\eta\to 0} \left(
\frac{1}{\eta^2}-\frac{J\sqrt{1-F^2}}{\eta F}\right) < \infty \ , 
\label{eq:boundsGJH}
\eeq
where we have used the expression for $H$ in Eq.~(\ref{GHI}).  From
here we learn that $G(0)=J(0)$, $H(0)=0$ and also this yields a
refinement of the boundary condition (\ref{JsuF}) 
\beq
\lim_{\eta\to 0} \frac{J}{F} = \lim_{\eta\to 0} \frac{1}{\eta}
+ \mathcal{O}(\eta) \ . \label{JsuFdue}
\eeq
If we now expand $F$ as
\beq
F \sim c_j \eta^j + \mathcal{O}\left(\eta^{j+1}\right) \ , 
\eeq
where $j=1,2,3,\ldots$ and $j$ represents the first non-zero term in
the expansion then Eq.~\eqref{JsuFdue} implies
\beq
J \sim c_j \eta^{j-1} + \mathcal{O}\left(\eta^{j+1}\right)
\ . \label{eq:Jexpansion}
\eeq
This is sufficient for rendering $H/\eta$ finite at small $\eta$.
Finally, $G \sim j c_j \eta^{j-1} + \cdots$ together with
Eq.~\eqref{eq:Jexpansion} implies that the first constraint of
Eq.~\eqref{eq:boundsGJH} is trivially satisfied for $j>1$ while it 
is also non-trivially satisfied for $j=1$ due to the matching of the 
coefficients of $G$ and $J$.

The other components of the field strength can be evaluated easily.
Using the fact that $I=\eta G$ from Eq.~(\ref{GHI}) one finds the $y$
component of the magnetic field 
\beq
F_{13}|_{y=0} = -\frac{\alpha\p}{z}G \, t^2 \ ,
\eeq
which yields a finite energy contribution at large $z$ since
$\alpha\p$ is at most of order $\mathcal{O}(1/z^2)$.  A divergence is
avoided at small $z$ if in addition to (\ref{alphazz}) one imposes 
\beq
\lim_{z\to 0}\frac{\alpha\p}{z} < \infty \ . 
\eeq
In fact this condition follows from (\ref{alphazz}) if $\alpha$ is
differentiable at $\eta=0$. 

The final component of the magnetic field is 
\beq
F_{23}|_{y=0} = 
\left[\frac{\alpha \eta}{z^2}
  \left(\frac{G - J}{\eta}
  - \frac{\alpha G H}{n} - J'\right) 
  + \frac{\alpha' J}{z}\right]t^1
- \left[\frac{\alpha \eta}{z^2}
  \left(-\frac{H}{\eta} + \frac{\alpha J G}{n} - H'\right) 
  + \frac{\alpha' H}{z}\right]t^3 \ .
\eeq
Again the finiteness of this contribution to the energy is guaranteed
by the condition Eq.~(\ref{alphazz}).

\subsection{What lies outside of the cones}\label{sec:outsidethecones}

To choose a consistent configuration outside of the cones, one must
first determine the configuration on the boundaries of the cones.  We
have already imposed the boundary condition $F(\sigma)=0$ and we have
seen that the finiteness of $H(\sigma)$ implies that $J(\sigma)=0$ as
well.  In general the shape of the region between the cones is quite
complicated.  However, we are only interested in an approximately
BPS configuration, which we define as a series of configurations at 
various values of $Q$ such that at large $Q$ the energy is
asymptotically proportional to $Q$.  We will see in this subsection
that it is therefore sufficient to consider a configuration in which
the gauge field vanishes in the region between the cones, considerably
simplifying our analysis. 

The vanishing of the gauge field in the region between the cones means
that it also vanishes on the boundaries of the cones 
\beq
G(\sigma) = H(\sigma) = I(\sigma)=0 \ .
\eeq
According to Eq.~(\ref{GHI}), for $G(\sigma)$ and $I(\sigma)$ to
vanish it is sufficient to fix 
\beq
F\p(\sigma) = 0 \ ,
\eeq
which is the reason for the modification proposed in
Eq.~\eqref{eq:Fmodsine}.
The vanishing of $H(\sigma)$ is only slightly more complicated.  By
(\ref{GHI}), now with $\epsilon=-1$, 
\beq
- \frac{1}{\sigma} = 
\lim_{\eta\to\sigma} \frac{J(\eta)\sqrt{1-F^2(\sigma)}}{F(\eta)} \ .
\eeq
Since $F\p(\sigma)$ vanishes, so must $J\p(\sigma)$ and so taking yet
another derivative of the numerator and denominator 
\beq
J'(\sigma) = 0 \ , \qquad 
J''(\sigma) = - \frac{F''(\sigma)}{\sigma} \ . \label{jf}
\eeq

With these choices the gauge field vanishes on the boundary and the
Higgs field is equal to 
\beq
\Phi = - h(z)\, t^3 = -h\left(r/\sqrt{1+\sigma^2/n^2}\right)t^3 \ . 
\label{phifuori}
\eeq
These fields may then be easily extended to the region outside of the
cones, simply by asserting that $A_i$ always vanishes and $\Phi$
always obeys (\ref{phifuori}).  This choice certainly does not minimize the
energy, but as we will now argue, it yields a contribution to the
energy which at large $Q$ is increasingly subdominant. 

Notice that the energy density between the cones is simply
$|\partial_i \Phi|^2$ which depends only upon $r$.  Thus it is equal
to a $Q$-dependent constant multiplied by the area of a sphere of
radius $1$ which is not within a cone.  This area tends to a constant
at large $Q$.  In fact, at large $Q$ the maximal number of cones that
can be packed into a finite volume approaches the maximum packing of
their circular cross-sections on a plane, which is given by the ratio
of the area of a unit hexagon to a unit circle $2\sqrt{3}/\pi$.  The
area at unit radius outside of a cone is then $8\sqrt{3}-4\pi$ and
hence, integrating over the radius
\beq
E_{\rm out} = \frac{\left(8\sqrt{3}-4\pi\right)}{g^2}
\int dr \, r^2 \Tr\left|\partial_i \Phi\right|^2 = 
\frac{\left(8\sqrt{3}-4\pi\right)}{g^2} 
\int dr \, r^2 \left(h'(r)\right)^2 \ , 
\eeq
yields the energy outside of the cones.

Therefore at large $Q$ the geometric factor asymptotes to a
$Q$-independent constant.  Also at large $Q$ the $\sigma/n$ in the
argument in Eq.~(\ref{phifuori}) becomes negligible.  Therefore the
only source of $Q$-dependence is in $h$ itself, which interpolates
between $0$ and the $Q$-independent constant $gv$.  We will argue in
Sec.~\ref{qualsez} that this interpolation occurs over a region of
size of order $\mathcal{O}(\sqrt{Q})$.  Therefore one expects that the
derivatives will decrease, and in particular the derivative squared
energy density will scale as $1/Q$.  However the range of integration
scales as $\sqrt{Q}$, and so one expects the total contribution to the
energy between the cones to decease as $1/\sqrt{Q}$, becoming ever
subdominant as compared with the BPS energy which is proportional to
$Q$ itself.

\subsection{The total energy and BPS equations}

At an arbitrary point in space, the covariant derivatives of $\Phi$ are
\begin{align}
D_1\Phi &= 
\left(c^2 X_1 + s^2 X_2\right) t^1 
+ cs \left(X_1 - X_2\right) t^2
- c X_3 t^3 \ , \\ 
D_2\Phi &= 
cs \left(X_1 - X_2\right) t^1
+ \left(s^2 X_1 + c^2 X_2\right) t^2 
- s X_3 t^3 \ , \\
D_3\Phi &= 
\left(Z_1 - \frac{\eta}{n} X_1\right) \left(c t^1 + s t^2\right) 
+ \left(Z_2 + \frac{\eta}{n} X_3 \right) t^3 \ , 
\end{align}
where we have defined the following functions
\begin{align}
X_1 &\equiv \frac{h F'}{z}\left(n-\alpha\right) \ , &\quad
X_2 &\equiv \frac{h F}{z\eta}\left(n-\alpha\right) \ , \qquad
X_3 \equiv \frac{h F G}{z}\left(n-\alpha\right) \ , \\
Z_1 &\equiv h' F \ , &\quad
Z_2 &\equiv h' \epsilon \sqrt{1-F^2} \ , 
\end{align}
and the field strength components are
\begin{align}
F_{12} &= 
\frac{n}{\eta} U_1 \left(c t^1 + s t^2\right)
  + \frac{n}{\eta} U_2 t^3 \ , \\
F_{23} &= 
\left[c^2\left(W J + U_1\right) + s^2 W G\right] t^1 
  + cs\left(W(J-G) + U_1\right) t^2
  + c\left(U_2 - W H\right) t^3 \ , \\
F_{31} &= 
cs\left(W(J-G) + U_1\right) t^1
  + \left[c^2 W G + s^2\left(W J + U_1\right)\right] t^2 
  + s\left(U_2 - W H\right) t^3 \ , 
\end{align}
where we have defined
\begin{align}
U_1 \equiv \frac{\alpha}{z^2}\left[
G - J - \eta\left(J' + \frac{\alpha}{n} G H\right)\right] \ , \quad
U_2 \equiv \frac{\alpha}{z^2}\left[
H + \eta H' - \frac{\alpha}{n} \eta G J\right] \ , \quad
W &\equiv \frac{\alpha'}{z} \ .
\end{align}

The BPS equations read
\begin{align}
Z_1 - \frac{\eta}{n} X_1 - \frac{n}{\eta} U_1 = 0 \ , \qquad
Z_2 + \frac{\eta}{n} X_3 - \frac{n}{\eta} U_2 = 0 \ , \label{eq:BPS}\\
X_1 = W J + U_1 \ , \qquad
X_2 = W G \ , \qquad
X_3 = W H - U_2 \ . \nonumber
\end{align}
while the energy density for BPS-saturated configurations reads
\begin{align}
g^2 \mathcal{H}_{\rm cone,BPS} &= 
\epsilon^{ijk}\partial_i\Tr\left(F_{jk} \Phi\right) = 
\epsilon^{ijk} \Tr\left(F_{ij} D_k \Phi\right) \non &=
\frac{n}{\eta} \left(U_1 Z_1 + U_2 Z_2\right)
+ W \left(J X_1 + G X_2 + H X_3\right) \ . 
\end{align}
The only function of $\eta$ which is not given in terms of the
function $F$ is $J$. 

We will in the following formally integrate the
above boundary term, obtaining the correct magnetic
charge. This procedure imposes some criteria on $J$. Using the relations
\eqref{GHI} we obtain the topological contribution to the energy density
\begin{align}
g^2\mathcal{H}_{\rm cone,BPS} = 
\frac{\alpha h'}{z^2\eta}\left[
(n-\alpha)\frac{\epsilon F F'}{\sqrt{1-F^2}}
+\frac{n\eta J F'}{F^2}
-\frac{n\left(J+\eta J'\right)}{F}\right] 
+\frac{\alpha' h}{z^2\eta} 2(n-\alpha)
  \frac{\epsilon F F'}{\sqrt{1-F^2}} \  . \label{eq:topoen}
\end{align}
The total energy of the cone is given by
\beq
E_{\rm cone} = \frac{2\pi}{n^2}\int dz\int_0^\sigma d\eta\; z^2 \eta \,
\mathcal{H}_{\rm cone} \ , \label{eq:Econe}
\eeq
which tells us that it will be convenient to have the energy density
of the form $\frac{1}{\eta}\frac{d(\cdots)}{d\eta}$. After
this integration,  the BPS part of the energy depends only on the boundary data and so by Stokes' theorem is of the form
$\frac{1}{z^2}\frac{d(\cdots)}{dz}$. Rewriting Eq.~\eqref{eq:topoen}
as discussed
\begin{align}
g^2\mathcal{H}_{\rm cone,BPS} = &\  
\frac{\alpha h'}{z^2\eta}\left[
(n-\alpha) \partial_\eta\left(-\epsilon\sqrt{1-F^2}\right)
-n \partial_\eta\left(\frac{\eta J}{F}\right)\right] \non&
+\frac{\alpha' h}{z^2\eta} 2(n-\alpha)
  \partial_\eta\left(-\epsilon\sqrt{1-F^2}\right) \ ,
\end{align}
we see from the first and third terms that
$-\epsilon\sqrt{1-F^2}|_0^{\sigma}=2$. Hence, in order to obtain a total
derivative in $z$, we need to impose
\beq
\left.-\frac{\eta J}{F}\right|_0^{\sigma} = 2 \ . 
\eeq
This can be done consistently with the boundary condition (\ref{jf}) in many ways, among which we will choose
\beq
J = \frac{F}{\eta}\epsilon\sqrt{1-F^2} \ . \label{eq:Jsuggestion}
\eeq 
Carrying out first the $\eta$ integration, we obtain
\begin{align}
\int_0^{\sigma} d\eta \; \eta \, \mathcal{H}_{\rm cone,BPS} = 
\frac{2\alpha h'}{g^2 z^2} (2n-\alpha) 
+\frac{2\alpha' h}{g^2 z^2} 2(n-\alpha) =
\frac{2}{g^2 z^2}\partial_z\left[(2n-\alpha)h\alpha\right] \ ,
\end{align}
which can readily be integrated 
\beq
E_{\rm cone,BPS} = \frac{4\pi}{g^2 n^2}
\left[(2n-\alpha)h\alpha\right]_0^{\infty}
= \frac{4\pi v}{g} \ . \label{eq:BPSenergy}
\eeq
This is exactly one Dirac quantum contained in a single cone. 

We will however see that we can only approximately satisfy the above
BPS equations and hence we will need to calculate the total energy
density 
\begin{align}
\mathcal{H}_{\rm cone} = \frac{1}{2g^2}\bigg[&\,
X_1^2 + X_2^2 + X_3^2
+\left(Z_1 - \frac{\eta}{n}X_1\right)^2
+\left(Z_2 + \frac{\eta}{n}X_3\right)^2 \label{eq:coneenergy} \\&
+\left(1+\frac{n^2}{\eta^2}\right)\left(U_1^2 + U_2^2\right)
+2 W \left(J U_1 - H U_2\right)
+W^2\left(G^2 + H^2 + J^2\right) \bigg] \ , \nonumber
\end{align}
for our approximate solutions. The first five terms are the
kinetic energy contributions given in
Eq.~\eqref{eq:rhokin}.

The total energy of the monopole is then
\beq
E_{\rm monopole} = Q E_{\rm cone} + E_{\rm out} \ . \label{eq:Etotal}
\eeq
Ideally the BPS equations could all be satisfied, in which case
$E_{\rm cone}=E_{\rm cone,BPS}$, however, as we will see this does not
turn out to be the case for our configuration. Hence the energy of the
cone is given by Eq.~\eqref{eq:Econe} with $\mathcal{H}_{\rm cone}$
given by Eq.~\eqref{eq:coneenergy}. This energy will necessarily
exceed the BPS bound \eqref{eq:BPSenergy}.

\subsection{Radial profile}

The conditions in Secs.~\ref{sec:finiteenergyconditions} and
\ref{sec:outsidethecones} guarantee that the energy of these
configurations is finite.  However, a qualitative agreement with true
BPS solutions is only possible if the total energy is, at large $Q$,
proportional to $Q$.  Such a scaling constrains the radial dependence
of the Higgs field and gauge fields.  There are many different choices
of radial dependence which yield the correct scaling\footnote{Indeed two such choices were described yesterday in Ref.~\cite{Manton:2011vm}.}.  We will choose
one of the simplest, we will impose that the Bogomol'nyi equations be
satisfied near the cores of the cones.  More precisely, we will expand
the solution as a power series in $\eta/n$ and will apply the
Bogomol'nyi equations at leading order in the expansion. This will
determine the radial profile functions. 

If we expand the function $F$ as follows
\beq
F(\eta) = a \eta - \frac{1}{6}b\eta^3 + \mathcal{O}(\eta^5) \ , 
\eeq
and choose 
\beq
J = \frac{F}{\eta}\epsilon\sqrt{1-F^2} \ ,
\eeq
as was suggested in Eq.~\eqref{eq:Jsuggestion}, 
we obtain at leading order in $\eta/n$ the following ODEs 
\begin{align}
h(z)  &= \frac{\alpha'(z)}{n-\alpha(z)} \ , \label{eq:h} \\ 
h'(z) &= \frac{a^2 \alpha(z)}{z^2}(2n - \alpha(z)) \ , \label{eq:hp}
\end{align}
which can be combined into a single equation for $\alpha$
\beq
\frac{\alpha''(z)}{n-\alpha(z)}
+\left(\frac{\alpha'(z)}{n-\alpha(z)}\right)^2
-\frac{a^2 \alpha(z)}{z^2}(2n - \alpha(z)) = 0 \ . \label{eq:diffeq}
\eeq
With the radial differential equation at hand we are now ready to
write down all of the angular profile functions in the next subsection.

\subsection{Choosing an angular profile function}

Summarizing the construction of this section, given the functions 
$F$ and $J$ one may determine all of the angular functions in the Ansatz 
(\ref{eq:phi},\ref{eq:A}). 
The radial functions on the other hand are determined by the
Bogomol'nyi equations (\ref{eq:h}) and (\ref{eq:hp}) at
leading order in $\eta/n$. 

For instance, if $F$ is given by Eq.~\eqref{eq:Fmodsine} and $J$ by
Eq.~\eqref{eq:Jsuggestion}, then the other angular functions are given by 
\begin{align}
G(\eta) \simeq 
\frac{\left|\frac{\pi}{\sigma}\cos\left(\frac{\pi\eta}{\sigma}\right)
\left(1-\frac{\eta^\kappa}{\sigma^\kappa}\right)
-\frac{\kappa\eta^{\kappa-1}}{\sigma^\kappa}
\sin\left(\frac{\pi\eta}{\sigma}\right)\right|}
{\sqrt{1-\sin^2\left(\frac{\pi\eta}{\sigma}\right)
  \left(1-\frac{\eta^\kappa}{\sigma^\kappa}\right)^2}} \ , \quad
I(\eta) = \eta \, G(\eta) \ , \quad
H(\eta) = \frac{F^2(\eta)}{\eta}  \ , 
\end{align}
whose $\eta/n$ expansion contains
\eqref{eq:diffeq} with
\beq
a = \frac{\pi}{\sigma} \ . 
\eeq

As we have mentioned, we will not be able to satisfy the BPS
equations everywhere (or equivalently to all orders in the $\eta/n$ expansion) due to the fact
that true BPS solutions do not exactly factorize. 
Hence, in order to measure the excess energy with respect to a true
BPS configuration we need to calculate the total energy density as
follows 
\begin{align}
g^2\mathcal{H}_{\rm cone} = & \,
\frac{1}{2} h'(z)^2
+\left(\frac{\sigma^2}{n^2} \mathcal{P}(\eta,\sigma)
+\mathcal{Q}(\eta,\sigma)\right)
\frac{h(z)^2\left(n-\alpha(z)\right)^2}{z^2} \\&
+\mathcal{Q}(\eta,\sigma)\frac{\alpha'(z)^2}{z^2}
+\left(\frac{1}{\sigma^2}\mathcal{R}(\eta,\sigma)
+\frac{1}{n^2}\mathcal{S}(\eta,\sigma)\right)
\frac{\alpha(z)^2\left(2n-\alpha(z)\right)^2}{z^4} \ , \nonumber
\end{align}
which should be integrated over the volume of the cone as in
Eq.~\eqref{eq:Econe} while the total energy is given by
Eq.~\eqref{eq:Etotal}. All the above functions $\mathcal{P}$,
$\mathcal{Q}$, $\mathcal{R}$ and $\mathcal{S}$ are defined such that
their integral $\int_0^\sigma d\eta \; \eta X = {\rm const}$ is
independent of $\sigma$, where $X=\mathcal{P}$,
$\mathcal{Q}$, $\mathcal{R}$, $\mathcal{S}$.

By evaluating the integrals over the angular functions numerically we
can write the total energy for the cone
\begin{align}
E_{\rm cone} \simeq \frac{2\pi}{g^2 n^2}\int dz \; z^2 \bigg[ & \,
\frac{1}{4} \sigma^2 h'(z)^2
+\left(1.24 \frac{\sigma^2}{n^2}+3.11\right)
\frac{h(z)^2\left(n-\alpha(z)\right)^2}{z^2} \\&
+3.11\frac{\alpha'(z)^2}{z^2}
+\left(\frac{6.02}{\sigma^2}
+\frac{1.24}{n^2}\right) 
\frac{\alpha(z)^2\left(2n-\alpha(z)\right)^2}{z^4} \bigg]
\ . \nonumber 
\end{align}
One learns from this expression that there is a competition in energy
which is determined by the value of $\sigma$. Only the first and the
last term are dependent on $\sigma$ at large $n$ and the first wants
$\sigma$ to be small while the last one prefers a large value of
$\sigma$. It so happens that the last term wins the competition due
to the fact that $h$ varies only over a fairly small fraction of the
integration range while the last term is roughly $\frac{6
  n^4}{\sigma^2 z^2}$ which is large for large $n$. Hence a
minimization of energy leads to a value of $\sigma$ which is as large as possible,
i.e.~$\sigma\sim 1.9$ as found in Eq.~\eqref{eq:sigmamax}. We will
check this statement numerically in Sec.~\ref{numsez}.

\section{Asymptotics and qualitative features} \label{qualsez}

As we have described, the monopole is characterized by the
configuration in a single cone.  This consists of two parts, an angular
function $F$ which needs to be chosen as well as radial functions
$\alpha$, $h$. 
The factorizing Ansatz (\ref{eq:phi},\ref{eq:A}) implies that the
radial functions are independent of the choice of $F$.  In
particular the function $\alpha$ may be determined using the second
order ODE (\ref{eq:diffeq}) and then $h$ may be determined from
$\alpha$ using Eq.~(\ref{eq:h}).  

\subsection{Small $z$ asymptotic behavior}

At small $z$, near the center of the monopole, the gauge and Higgs 
fields approach zero. These two conditions imply the boundary
conditions 
\beq
\alpha(0) = \alpha'(0)=0 \ . \label{eq:alpha_bordozero}
\eeq
In the small $z$ region, $\alpha,\alpha'\ll n$. 
Therefore (\ref{eq:diffeq}) may be approximated by
\beq
\alpha''(z) - \frac{2 a^2 n^2\alpha(z)}{z^2} = 0 \ . \label{aeqb}
\eeq
This is a homogeneous, linear, ODE with the solution 
\beq
\alpha(z) = k_1 z^{e_1} \ , \qquad
e_1 = \frac{1}{2} + \sqrt{\frac{1}{4} + 2a^2n^2} \ .
\label{b1} 
\eeq
At large $n$, corresponding to large $Q$, this implies 
\beq
\alpha(z) \simeq k_1 z^{\sqrt{2}a n} \ . \label{apic}
\eeq
The function $h$ can then be found from the small $z$ limit of
Eq.~\eqref{eq:h} 
\beq
h(z) \simeq \frac{\alpha'(z)}{n} = \sqrt{2}a
  k_1 z^{\sqrt{2}a n - 1} \ . 
\eeq

\subsection{Large $z$ asymptotic behavior}

Far from the monopole, a finite energy solution requires that $\alpha$
tends to $n$.  It will prove convenient to define the function 
\beq
\beta(z) \equiv n - \alpha(z) \ , 
\eeq
which tends to $0$ at large $z$.  Eq.~\eqref{eq:diffeq} may
be re-expressed in terms of $\beta$ as 
\beq
-\frac{\beta''(z)}{\beta(z)}
+\left(\frac{\beta'(z)}{\beta(z)}\right)^2
+\frac{a^2}{z^2}\left(\beta(z)^2-n^2\right) = 0 \ . \label{beq} 
\eeq
We will now assume that $\beta$ may be expanded at large $z$ such that 
the leading term is 
\beq
\beta(z) = k_2 z^{e_2} e^{-m z} \label{eq:banz}
\eeq
where $m>0$.  
Substituting this into (\ref{beq}) one finds
\beq
\frac{e_2-a^2 n^2}{z^2} + \cdots = 0 \ ,
\eeq
where the ellipsis denote exponentially suppressed terms, which may be
canceled by subdominant corrections to (\ref{eq:banz}).  Therefore we
demand only that the first term vanishes, fixing $e_2=a^2n^2$.

Again the Higgs field profile function $h$ is easily found from
(\ref{eq:h}) 
\beq
h(z) = - \frac{\beta'(z)}{\beta(z)}
     = m - \frac{a^2 n^2}{z} \ . \label{eq:hgrande}
\eeq
BPS monopoles are characterized by the boundary condition that $h$
tends asymptotically to the vacuum expectation value $g v$ (recall 
that we have rescaled the coupling into the scalar field).  Therefore
$m$ is determined entirely by this boundary condition 
\beq
m = g v \ .
\eeq
Physically, $m$ is the mass of the $W$-bosons, and the exponential decay
in Eq.~(\ref{eq:banz}) is just that of a massive field.  Our final
asymptotic form for the function $\beta(z)$ is hence
\beq
\beta(z) \simeq k_2 z^{a^2 n^2} e^{-g v z} \ . \label{eq:bgrande}
\eeq

\subsection{Connecting the regimes and the monopole bag}\label{sec:conn}

We have studied two regimes:  one, at small $z$, where $\alpha\ll n$
and another, at large $z$, where $\beta=n-\alpha\ll n$.  Clearly it is 
important to determine at what value $z=z_0$ the solution of
(\ref{eq:diffeq}) interpolates between these two regimes.  Neither 
approximation of $z$ leads to a well-controlled expansion in the
intermediate regime.  However one may arrive at a qualitative
understanding of the solution by imagining that both expansions are
roughly correct, in the sense that will be described below, at $z_0$.
This is potentially a dangerous assumption, but numerically we have
verified that the results of this subsection are indeed correct. 

We will define $z_0$ to be the midpoint of the function $\alpha$
\beq
\alpha(z_0) = \beta(z_0) = \frac{n}{2} \ .
\eeq
Therefore the matching of the two limits yields
\beq
\frac{\alpha'(z_0)}{\alpha(z_0)} = 
  -\frac{\beta'(z_0)}{\beta(z_0)} \ . 
\eeq
This relation is exact.  But the consequences may be approximated by
substituting the asymptotic behaviors (\ref{apic}) of $\alpha$ at
small $z$ and (\ref{eq:bgrande}) of $\beta$ at large $z$ which gives
an estimate for 
\beq
z_0 \sim \frac{a n}{g v} \left(\sqrt{2} + a n\right) 
  \sim \frac{a^2 n^2}{g v} \ , 
\label{eq:z0}
\eeq
where in the rightmost expression we have approximated $n\gg 1$. From
this expression it is clear that the $1/z$ tail of the Higgs field
gives the monopole its size $\propto n^2$.

The value of $z_0$ in Eq.~(\ref{eq:z0}) is the approximate size of the
monopole, the value of the radius at which the $W$-bosons begin to
fall exponentially.  At charge $1$, corresponding to $n=1$, one
recovers the fact that the monopole size is the inverse $W$-boson
mass.  However more generally it reveals, as was conjectured in
Ref.~\cite{Bolognesi:2005rk}, that the radius of the monopole is
proportional to the charge $Q=n^2$.  

One may also determine the thickness $w$ of the boundary region, the
distance over which $\alpha$ tends from a value near $0$ to a value
near $n$.   This depends on how quickly $\alpha$ changes at the
boundary $z_0$ 
\beq
\alpha'(z_0) = \sqrt{2}a n \frac{\alpha(z_0)}{z_0}
  \simeq \frac{g v}{\sqrt{2}a}
\eeq
valid for $n\gg 1$. 
The distance over which $\alpha$ changes by $n$ units is then approximately
\beq
w \sim \frac{n}{\alpha'(z_0)} \sim 
 \frac{\sqrt{2}a n}{g v} \ . 
\eeq
Therefore the width of the boundary of the monopole is proportional to
$n=\sqrt{Q}$, while the radius of the monopole is proportional to
$n^2=Q$.  Thus at large $n$ the walls of the monopole are much
thinner than its radius, confirming a conjectured description of these
solutions as monopole bags in Ref.~\cite{Bolognesi:2005rk}.  However
one should note that, given the large moduli space of solutions, it
seems possible that there are other sequences of BPS solutions with
approximate spherical symmetry in the large $Q$ limit but with
different qualitative radial profiles, and so the bag description may
not apply to them. 

The boundary of the monopole is not only relatively thin, with a
width $w$ much smaller than $z_0$ at large $n$, but also it is very
sharp, as $\beta$ exponentially decays at large $z$ as seen from
Eq.~\eqref{eq:bgrande}.  As we have mentioned, this is due to the fact 
that the gauge field is massive.  On the other hand, the massless
Higgs field decays only as $\mathcal{O}(1/z)$, as seen in
Eq.~\eqref{eq:hgrande}.  Therefore the Higgs field does not 
exhibit such a sharp transition at the boundary of the monopole.

\section{Numerical results} \label{numsez}

\begin{figure}[!htp]
\begin{center}
\includegraphics[width=0.85\linewidth]{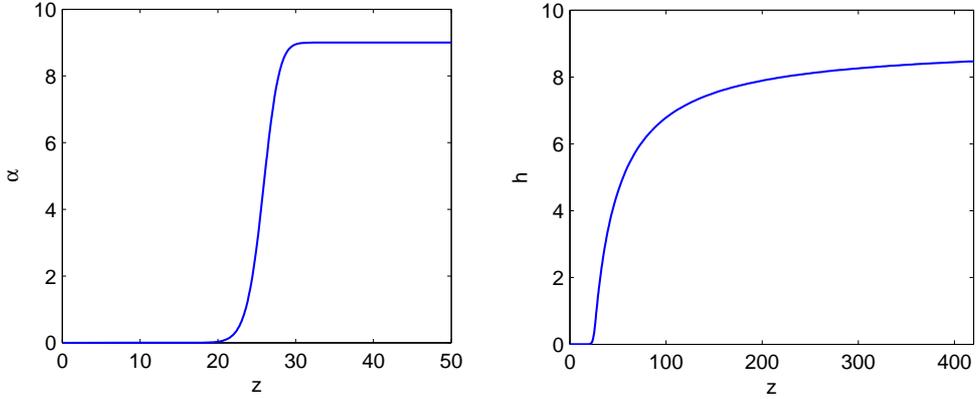}
\caption{Left panel: the profile function $\alpha$. Right panel: the
  profile function $h$ both as functions of $z$ for
  $Q=81,n=gv=9,\sigma=1.9,z_0=25.8$ and $a=\pi/\sigma$. }
\label{fig:n9}
\end{center}
\end{figure}

\begin{figure}[!htp]
\begin{center}
\includegraphics[width=0.85\linewidth]{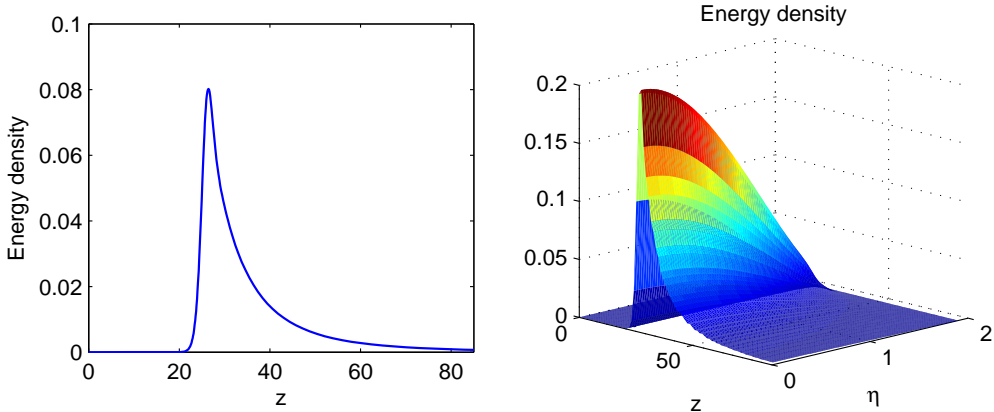}
\caption{Left panel: the radial energy density 
  $\int_0^\sigma d\eta \; \eta g^2\mathcal{H}_{\rm cone}$ as
  function of $z$ for the monopole with
  $Q=81,n=vg=9,\sigma=1.9$. Right panel: the energy density
  $g^2\mathcal{H}_{\rm cone}$ of the cone.}
\label{fig:n9energy}
\end{center}
\end{figure}

In this section we will provide a few numerical solutions to the ODE
\eqref{eq:diffeq} illustrating the radial profile functions for the
monopole. We have used a shooting method to find the solution using
Eq.~\eqref{apic} as the initial condition and hence $k_1$ as the
shooting parameter. A solution with $Q=81,n=gv=9,\sigma=1.9$ is shown
in Fig.~\ref{fig:n9} and its energy density is shown in
Fig.~\ref{fig:n9energy}. This solution has a total energy which is
$54\%$ higher than the BPS bound $4\pi v n^2/g$. 

\begin{figure}[!htp]
\begin{center}
\includegraphics[width=0.5\linewidth]{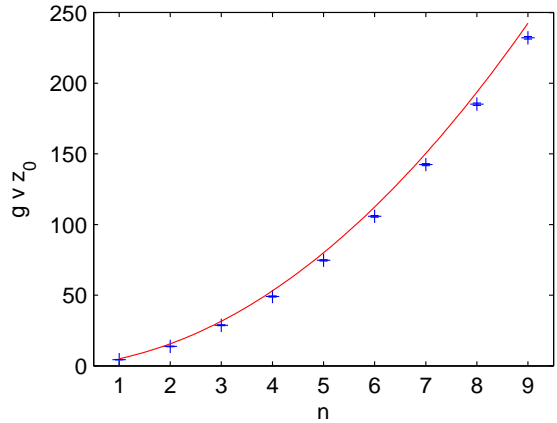}
\caption{The approximate monopole size $g v z_0$ which is
  predicted to be
  $\frac{\pi}{\sigma}n\left(\sqrt{2}+\frac{\pi}{\sigma}n\right)$
  and shown as the red line (we have set $\sigma=1.9$). The numerical
  results are shown with their corresponding errorbars. }
\label{fig:msize}
\end{center}
\end{figure}

In order to check the size of the monopole estimated in
Eq.~\eqref{eq:z0}, we have numerically calculated the value of $z_0$ such that $\alpha(z_0)=n/2$ as shown in Fig.~\ref{fig:msize}.

\begin{figure}[!htp]
\begin{center}
\includegraphics[width=0.5\linewidth]{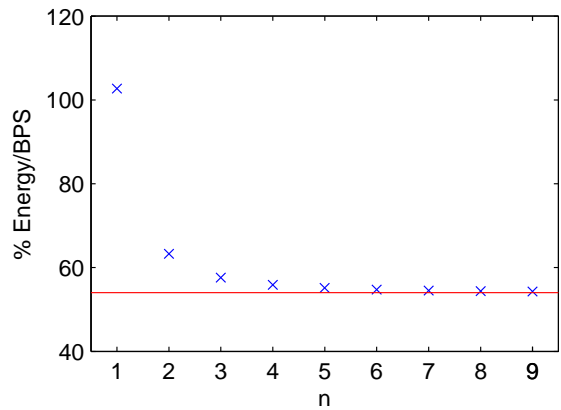}
\caption{The total excess energy of the monopole compared to the
  BPS-saturated energy percentage as a function of $n=\sqrt{Q}$. It is
  seen that the energy excess is roughly constant at large $n$ around
  $54\%$. }
\label{fig:nonbps}
\end{center}
\end{figure}

Finally we have checked the excess energy of the solutions compared
to the BPS-saturated ones which is shown in Fig.~\ref{fig:nonbps}. The
solutions turn out to exceed the BPS bound by roughly $54\%$ at large
$n$. 

\section{Future directions} \label{scurasez}

First of all we should provide a word of caution due to the fact that the
BPS solutions do not share the factorization property of our
configurations. For this reason 
we have calculated the radial profiles in terms of angular expansion
parameters expanded around the origin. This means that our monopole
configurations are only approximately BPS, and in particular do not
provide time-independent solutions of the equations of motion. However
we do believe they capture some 
key features of those true BPS monopoles which are spherically
symmetric. As a support to this claim we found numerically that at
high $Q$ the total energies of our configurations exceed that of true
BPS solutions by a $Q$-independent factor of roughly $54\%$.

As we have mentioned, it may seem as though we are studying a particularly
difficult and uninteresting part of the moduli space of
solutions. We would however like to conjecture that in a certain class
of theories it is the most interesting part:

\begin{conjecture}
The approximately spherically symmetric BPS monopoles are the only ones
which survive the strong non-BPS deformation described below.  They
all reduce to the same non-BPS configuration. 
\end{conjecture}

Our deep interest lies in non-BPS monopoles in which a Higgs potential
is included for the scalar $\Phi$.  We are interested in these
monopoles because of a series of perhaps coincidental facts relating
$Q=1$ {\it{non-BPS}} monopoles with the dark matter halos of the many
minimal size dwarf spheroidal galaxies which have recently been
discovered in our local group, for example by the Sloan Digital Sky
Survey.  Some of the most striking similarities, which in fact are
shared by other dark matter dominated galaxies such as larger dwarf
and low surface brightness spiral galaxies, are as follows: 
\newcounter{outline}

\begin{list}{\arabic{outline})} {\usecounter{outline} \setlength{\leftmargin}{0cm}\setlength{\itemsep}{.2cm}}
\item Dark matter halos, like topological solitons, have a minimum
  mass.  For solitons this corresponds to the charge $Q=1$.  The
  lightest satellites of the Milky Way have masses of about $10^7\,
  M_\odot$ within 1,000 light years of their 
center \cite{Kap1000} and between $10^7$ and $10^8\, M_\odot$ within
2,000 light years \cite{Strigari2000}.  This observed minimum dark
matter mass leads to the dwarf galaxy problem:  For particle models of
dark matter, like WIMPs, consistency with this minimum mass is a
problem because simulations generally suggest 4 to 400 times more
dwarf galaxies in our local group than have been observed, most of
which should be much lighter than the minimum observed mass.  In fact,
only one dwarf galaxy (ComaBer \cite{Belokurov}) 
has been observed with a mass near $10^6\, M_\odot$, and it is very
elongated and irregular and appears to be in the process of being
ripped apart by tidal forces.   

The existence of a topological charge, equal to one, for these small
galaxies not only explains the fact that no small dwarf galaxies have
been seen,  but also the related fact that smaller dark matter bodies
cannot exist.  In this way one avoids the fatal gravitational lensing
constraints faced by other MACHO dark matter models.  Indeed the upper
limit of the range of radii of dark matter candidates excluded by
gravitational lensing is many orders of magnitude smaller than these
1000 light year solutions, and so these  monopoles are too large to be
excluded by the lensing bounds. 
\item At least in cases in which there is enough visible matter to
  determine the density profile, dark matter dominated galaxies and
  non-BPS monopoles have cores with relatively constant densities,
  with intermediate regions with $1/r^2$ densities and external
  regions with a faster radial fall-off.  For $Q=1$ non-BPS monopoles
  this $1/r^2$ intermediate region density profile is inevitable,
  unlike the the higher $Q$ BPS $1/r^2$ density profile in the
  Bolognesi galaxy bags of Ref.~\cite{Manton:2011vm} in which there
  are many inequivalent choices of $r$-dependence, such as solutions
  which the author called planets, {\it{etc.}}
\item The cores of these non-BPS solutions in many cases naturally
  contain black holes \cite{LNW,BFM,HKK,Bolognesi:2010xt}.  In the
  case of the galaxies, models often suggest that there has not been
  enough time to form the supermassive black holes known to inhabit
  most galactic cores, in addition there are even some claims of
  supermassive black holes without luminous galactic hosts.  These
  problems are both naturally explained if the black hole is an
  integral part of the monopole solution, as it is in many models.
  The gradual consumption of stars, gas and dark matter particles is,
  in this scenario, no longer the main mechanism driving supermassive
  black hole formation. 
\item The simplest model in which non-BPS 't Hooft-Polyakov monopoles
  exist is a Georgi-Glashow model\footnote{Here we are considering a
    new sector, these gauge symmetries have no obvious relation with
    standard model or GUT symmetries and we do not specify the charges
    of standard model particles under this new gauge symmetry.} with a
  simple Abelian Higgs quartic potential.  In this case, a $Q=1$
  non-Abelian monopole with the radius $r$ and mass $M$ of the
  smallest dwarf galaxies arises if the value of the Higgs VEV is
  about $v\sim\sqrt{\hbar c^3M/r}\sim 10^{14}\ {\rm GeV}$.  This number only
  changes by about a factor of 2 depending on whether the luminous
  region is within the core or the intermediate radius regime.   Had
  $v$  been above the Planck scale, gravity would have dominated over
  the Georgi-Glashow interactions and the whole solution would have
  been a hairy black hole instead of a dwarf galaxy.  Had it been
  smaller than about $100\ {\rm eV}$, these monopoles could not have formed
  in time for dark matter to have played its crucial role in the
  oscillations of primordial plasma which reproduces the oscillation
  spectrum observed in the CMB.  Given that the two physical inputs in
  this calculation are of galactic scales, the fact that the output is
  a particle physics scale in this relatively narrow acceptable window
  is for us miraculous.  If one naively uses the rotation curves of
  slightly larger dwarves one may similarly conclude that the Higgs
  coupling is of order $\lambda\sim 10^{-97}$, had it entered $v$ with
  a different power, even a fourth root, the relation with dwarf
  galaxies would have been ruined. 
\item Similarly the 1,000 light year scale radii of these solutions
  imply that they form when the universe is about 1,000 years old
  \cite{kibble}.  Again, this is in time to help increase the
  intensity of fluctuations in the primordial plasma as is required by
  observations of the CMB.  Had dwarf galactic radii been larger by a
  factor of 100, they would have formed too late and an inconsistency
  would have arisen. 
\item While the monopole core excluding gravity has a constant
  density, and with gravity may host a black hole, the core itself is
  nonsingular.  More precisely it avoids the cusp problem of the
  $\Lambda CDM$ model, in which many simulations predict galactic mass
  distribution profiles, such as the historic  Navarro, Frenk and
  White profile \cite{NFW}, with density cusps in their cores, in
  stark contrast with observations. 
\end{list}

While these similarities are very strong, there is a serious problem
with galaxy sized non-BPS Bolognesi bags as a dark matter candidate.
Non-BPS monopoles repel, and as $v$ is less than the Planck scale this
repulsion would dominate over gravity and all galaxies would be
minimal dwarves and would repel one another.  There is a similar
problem of course for visible matter, which is mostly made of protons
which also repel.  In the case of protons, the solution is quite
complicated.  First of all there are electrons which screen the
interactions between protons.  While electrons have antiparticles
which have the same charge as protons, for reasons which have not yet
been quantitatively explained by any model, there was a primordial
excess of negatively charges electrons and positively charged protons.
They did not annihilate each other because they carry different
conserved charges.  They can combine, forming hydrogen bound states or
via inverse beta decay they can even merge beyond recognition into
neutrons.  However, due to the choice of parameters in the standard
model, the later possibility is kinematically disfavored in the
conditions that have existed in most of the universe since
baryogenesis. 

We would like to propose that repulsion between non-BPS monopoles is
avoided in a similar manner.  Additional conserved charges are easily
introduced in a Georgi-Glashow model by including charged fermions,
which via the Jackiw-Rebbi mechanism provide an additional charge for
each kind of monopole.  If one adds two species of fermions, then
there are two kinds of charge, which can play a role analogous to
baryon number and lepton number.  Monopoles of different charges can
have very different masses, in fact in $\mathcal{N}=1$ supersymmetric
models some flavors are usually massless while some are massive.  One
can then demand that the dark matter halos are made of very massive 
magnetically positively charged monopoles which carry one kind of
flavor charge, and that the screening is caused by light negative
monopoles with the other flavor charge.  This eliminates the problem
of galaxies repelling another. 

But one still needs to worry about the stability of $Q>1$ monopoles.
These will be held together by gravity.  Due to the mass of the scalar
field $\Phi$, the repulsive $SU(2)$ dynamics will dominate over the 
attractive scalar dynamics at large distances, leading to a net
repulsion.  The gravitational interaction in general is insufficient
to counter this repulsion, as $v$ is much less than the Planck mass.
However at large distances one expects the screening to play a role.
Unfortunately the exact role played by this screening is highly model
dependent and is not clear whether there exist any models which screen
this repulsion sufficiently to allow it to be dominated by
gravitational attraction, without adding any new interactions (playing
the role of strong interactions in the proton analogy described
above).  Strong constraints on models also arise from the fact that
one does not want the light monopoles to combine with the heavy
monopoles, analogously to inverse beta decay, as the resulting bound
state may not share the attractive features of the massive monopoles
described above. 

Therefore the model-independent predictions for $Q>1$ monopoles are
limited by ambiguities in the screening mechanism.  Nonetheless a
number of very firm predictions can be made.  First of all, just as
the dwarf galaxy problem is a gap between the mass of globular
clusters at $Q=0$ and dwarf galaxies at $Q=1$, there must also be a
gap between $Q=1$ and $Q=2$.  This prediction is much more general
than the monopole dark matter proposal discussed in this section, but
extends to any topological soliton dark matter candidate which solves
the dwarf galaxy problem by identifying minimal spherical dwarf
galaxies with $Q=1$ solitons.  The masses of these galaxies are at
best known at the 100 percent level, and so with current data such a
gap cannot be verified.  However one may hope that radio surveys of
gas in our galactic neighborhood such as that which will be performed by the FAST
telescope starting in 5 years will be able to test this claim. 

Another model independent prediction is that, while the charge $Q$ is
determined by the flat part of the galactic rotation curve, the radius
of the core must be proportional to $\sqrt{Q}$ and, even more
surprisingly, the outer radius of the region with the flat rotation
curves must at large $Q$ be nearly $Q$-independent.  This
counter-intuitive prediction may already rule out these models, as it
requires spatial extents of dwarf galaxy dark matter halos to extend
far beyond their most distant stars, but it is necessary for the
convexity of the galactic mass as a function of $Q$, which in turn is
necessary to prevent these galaxies from exploding. 

\subsubsection*{Acknowledgments}

We are eternally grateful to Stefano Bolognesi and Malcolm Fairbairn
for many useful and enlightening discussions.   JE is supported by the
Chinese Academy of Sciences Fellowship for Young International
Scientists grant number 2010Y2JA01.  SBG is supported by the Golda
Meir Foundation Fund.

\end{document}